\newif\ifsubmode
\submodefalse

\newif\ifprintfig
\printfigtrue

\newif\ifemulate
\emulatetrue

\ifsubmode
  \documentclass[12pt,preprint]{aastex}
  \received{}
  \accepted{}
  \journalid{}{}
  \articleid{}{}
\else
   \documentclass{emulateapj}
   \submitted{{\it Submitted for publication in AJ}}
\fi

\shortauthors{Willman, et al.}
\shorttitle{Stellar Tails in a Distant Galactic Satellite}

\def\lesssim{\mathrel{\hbox{\rlap{\hbox{\lower4pt\hbox{$\sim$}}}\hbox{$<$}}}}
\def\gtrsim{\mathrel{\hbox{\rlap{\hbox{\lower4pt\hbox{$\sim$}}}\hbox{$>$}}}}

\begin{document}

\title{Willman 1 - A Galactic Satellite at 40 kpc With Multiple Stellar Tails}

 \author{Beth Willman\altaffilmark{1}, Morad Masjedi\altaffilmark{1},
 David W. Hogg\altaffilmark{1}, Julianne
 J. Dalcanton\altaffilmark{2,3}, David
 Martinez-Delgado\altaffilmark{4,5}, Michael Blanton\altaffilmark{1},
 Andrew A. West\altaffilmark{2,6}, Aaron Dotter\altaffilmark{7}, Brian
 Chaboyer\altaffilmark{7}}


 \altaffiltext{1}{New York University, Center for Cosmology and
 Particle Physics, 4 Washington Place, New York, NY 10003,
 beth.willman@nyu.edu, david.hogg@nyu.edu mm1330@nyu.edu,
 mb144@nyu.edu}

 \altaffiltext{2}{Department of Astronomy, University of Washington,
 Box 351580, Seattle, WA 98195, jd@astro.washington.edu}

\altaffiltext{3}{Alfred P. Sloan Research Fellow}
 
 \altaffiltext{4}{Instituto de Astrofisica de Andalucia (CSIC),
 Granada, Spain}
 
 \altaffiltext{5}{Instituto de Astrofisica de Canarias, E-38205 La
 Laguna, Tenerife, Canary Islands, Spain, ddelgado@iac.es}

 \altaffiltext{6}{Astronomy Department, University of California,
 Berkeley, CA 94720-3411, awest@astro.berkeley.edu}

 \altaffiltext{7}{Department of Physics \& Astronomy, Dartmouth
 College, Hanover, NH, 03755, aaron.l.dotter@dartmouth.edu,
 brian.chaboyer@dartmouth.edu}


\ifsubmode\else
  \ifemulate\else \clearpage \fi
\fi


\ifsubmode\else
  \ifemulate\else
     \baselineskip=14pt
  \fi
\fi

\begin{abstract}
SDSSJ1049+5103, commonly known as Willman 1, is an extremely
low-luminosity Milky Way companion whose properties are intermediate
between those of globular clusters and dwarf spheroidals. In this
paper, we present new, wide-field photometry extending 3 mag below the
main sequence turnoff. These data show that this object is old,
moderately metal-poor, has a distance of 38$\pm$7 kpc and a half-light
radius of r$_{1/2} = 21\pm$7 pc, confirming previous
estimates. Willman 1's revised luminosity is $M_V$ = -2.5 mag, which
is somewhat fainter than the previous estimate. These new data show
that the total spatial extent of Willman 1 exceeds its tidal radius
for a range of assumptions about its total mass and its orbit,
suggesting it is significantly affected by the tidal field of the
Milky Way.  The spatial distribution of Willman 1's main sequence
stars also shows prominent multi-directional stellar tails.  The tidal
interactions causing these tail features may explain the large
physical size of Willman 1 relative to low-luminosity globular
clusters. At a distance of 40 kpc, it is the most distant Galactic
object yet known to display prominent tails, and is the only distant
satellite to display multi-directional tails. Although we cannot at
present determine the cause of this unusual morphology, preliminary
comparisons between the morphology of Willman 1 and published
simulations suggest that it may be near the apocenter of its orbit, or
that it may have interacted with another halo object.  We find a
significant difference between the luminosity functions of stars in
the center and in the tails of Willman 1, strongly suggesting mass
segregation much like that seen in Palomar 5. Although Willman 1 has
more pronounced tidal tails than most confirmed Milky Way dwarf
galaxies, because of its very low stellar mass we cannot at present
rule out the possibility that it has a dark matter halo.

\end{abstract}


\ifemulate\else
   \clearpage
\fi

\section{Introduction}
 The destruction of globular clusters and dwarf galaxies is thought to
 be an integral part of the formation of the Galactic stellar halo
 \citep{searle78,ashman98,bullock05}.  This theory is supported by the
 observed presence of tidal features around both globular clusters
 (GCs) and dwarfs.  For example, \citet{grillmair95} and
 \citet{leon00} found evidence for tidal features in nearly all of the
 12 and 20 GCs in their respective samples. The extensively studied
 tails of Palomar 5 ($d=23$ kpc) extend over 20 degrees of the sky and
 contain more stellar mass than the cluster itself
 \citep{odenkirchen03,grillmair06b}. More recently, \citet{belokurov06} and
 \citet{grillmair06} have used SDSS data to trace tails of NGC 5466
 ($d=16$ kpc) over 4 and 45 degrees of sky, respectively.

 Like GCs, many of the 10 known Milky Way (MW) dSphs show signs of
 tidal stripping. They exhibit breaks in their light profiles that are
 characteristic of tidal stripping, however the interpretation of both
 the light and velocity profiles of nearby dwarfs has been
 controversial \citep{wilkinson04,munoz05}. Of these dSphs, only
 Sagittarius (Sgr; $d=28$ kpc) exhibits tidal tails as prominent as
 those seen in globular clusters.



Stellar tails not only provide clues to the formation of the stellar
halo, but they also constrain the shape of the MW's dark matter halo
and the amount of substructure within it. For example, the Sagittarius
tidal stream has been used to measure the shape of the Galactic halo,
although there is not yet agreement as to whether the shape is oblate
or prolate \citep{ibata01, helmi04, johnston05}.  The coldness and
morphology of tidal tails are affected by the amount of substructure
in the Galactic halo, although present comparisons between
observations and models of tidal tails have not yet produced strong
constraints \citep{ibata02,johnston02b,mayer02,majewski04b}.

Tidal tails can also be used to infer properties intrinsic to the
object being stripped.  For example, the more a massive Galactic
satellite is, the less strongly it is affected by the tidal field of
the Milky Way. The presence of tidal features can thus be used to
limit the mass-to-light ratios of stripped objects
\citep{oh95,moore96,ibata97}. The tidal tails of
Palomar 5 have provided particularly interesting insights into its
past and present properties. \citet{odenkirchen03} and
\citet{dehnen04} used its inferred mass loss rate to derive an
initial mass that is more than an order of magnitude greater than its
present mass. They also used numerical models to determine a life
expectancy for Pal 5 of $\sim$ 110 Myr. From the fact that Pal 5's
life expectancy is so short, they conclude it is possible that
many similar objects once populated the Milky Way's inner halo.


 
 In this paper, we take a detailed look at a low density, outer halo
 object that we will show shares similarities with Palomar
 5. \citet{willman05a}, hereafter known as Paper I, presented the
 discovery of SDSSJ1049+5103, an old, moderately metal-poor Milky Way
 companion.  The nature of this object, which is commonly referred to
 as Willman 1, is ambiguous. In Paper I, we showed that its luminosity
 and half-light radius place it on the intersection between the
 size-luminosity relations followed by globular clusters and by old
 dwarf galaxies.  In this paper, we present new wide-field, deep
 imaging data and make more robust estimates of Willman 1's basic
 properties. We also show that Willman 1 has prominent
 multi-directional tails, making it the most distant object observed
 to have such features.  We show it is highly probable that at least
 some of the tail features are due to tidal interactions with the
 Milky Way.  The purely photometric dataset presented in the paper is
 still insufficient to either confirm or exclude the possibility that
 Willman 1 formed inside of its own dark matter halo.

 The outline of the paper is as follows. In \S2, we describe the
 observations, data reduction, and completeness tests. In \S3 we
 present results, including the clear presence of stellar tails and
 evidence for mass segregation of stars. We discuss these results in
 \S4.

\section{Data}
We obtained wide-field imaging of Willman 1 with MOSA on the 4m at
Kitt Peak National Observatory on 2005 April 7 and 8. This wide-field
mosaic camera is composed of 8 2048$\times$4096 chips with 0.26$''$ pixels,
resulting in a 34$'$$\times$34$'$ field of view.  Ten 600s exposures were
taken through each of the SDSS $g$ and $r$ filters at 5 different
positions dithered by $\pm$ 30 arcsec and $\pm$ 52 arcsec, with seeing
varying between 1.2$''$ and 1.4$''$. Images were bias-subtracted using
a sigma-clipped mean of darks (images taken with the shutter closed)
and flats were made by taking a sigma-clipped mean of dome flats. The
astrometric world coordinate system for each of the eight MOSA chips
for each exposure were determined independently by comparison to the
USNO-B1.0 astrometric catalog.  The DAOPHOT II/ALLSTAR package
\citep{stetson94} was used to obtain photometry of the resolved
stars. Stellar sources were selected as those identified with $chi$
$<$ 2 and $-0.5 < sharp < 0.5$ in at least 2 of the 10 exposures in
each filter.  Stellar magnitudes were photometrically calibrated with
the SDSS stellar catalog. The calibration uncertainty varied with
exposure and chip number, but is an average of $\sim 0.013$
magnitudes. The apparent magnitudes were corrected for extinction
using the \citet{schlegel98} dust maps. The average E(g-r) along
the line of sight to Willman 1 is 0.014.


\subsection{Completeness Testing}
 To determine the completeness of the stellar catalog as a function of
 magnitude, we conducted extensive artificial star tests.  We
 simulated artificial stars with $22.6 < r < 24.8$ and $0.25 < g-r <
 0.65$, the same range as Willman 1's main sequence stars. We inserted
 a grid of 4500 stars at a time into each exposure of the camera chip
 that covered Willman 1 and processed the simulated data as described
 in \S2. We repeated this procedure 10 times to simulate a total of
 45,000 artificial stars.  Artificial stars are considered detected
 only if: 1) the recovered positions are within 0.5$''$ of the input
 positions, and 2) the recovered magnitude lies within the same 0.5
 magnitude bin as the input magnitude.  An artificial star coincident
 with a ``real'' star is only counted as detected if it is brighter
 than the real star.  With these requirements, we retrieve:
 $\sim$100$\%$ of stars with r $<$ 22.5, $\sim$97\% of stars with 22.5
 $< r < 23.0$, $\sim$88\% of stars with 23.0 $< r < 23.5$, $\sim$84\%
 of stars with 23.5 $< r < 24.0$, and $\sim$75\% of stars with 24.0 $<
 r < 24.5$.  The completeness drops precipitously for stars fainter
 than 24.5. We thus consider 24.5 to be the completeness limit of our
 data.

 Although the artificial stars have colors and magnitudes consistent
 with those of Willman 1's main sequence, we have not precisely
 modeled the expected color-magnitude distribution of stars with the
 (very uncertain) age, distance and metallicity of Willman 1.  This
 completeness testing is thus not optimized to correct the stellar
 luminosity function (LF) in an absolute sense, because it does not
 correctly account for the scattering of stars in and out of the
 color-magnitude bins used to create the LF in \S3.2. However, since
 this scattering is a second order effect and since our primary
 concern in \S3.5 will be the relative LFs of stars in the outer and
 inner regions of Willman 1, our estimate of the completeness is
 sufficient for our scientific goals.  Due to the low surface density
 of Willman 1, crowding does not play a role in the completeness of
 stars brighter than $r$=24.5. There are only 140 stars brighter than
 $r$=24.5 within $1.0'$ of Willman 1's center. This translates to a
 mean spacing of 9$''$, which is roughly 7$\times$ the psf. We used our
 artificial star test to compare the completeness in the field and in
 the very center ($r < 0.6'$) of Willman 1 and found that they were
 identical. This comparison confirms that the relative luminosity
 function presented in \S3.5 is accurate.

\section{Results} \label{sec:results}

Figure 1 shows the resulting color-magnitude diagram (CMD) of stars
within the central 2$'$ of Willman 1. This region roughly corresponds
to the half-light radius, as derived in Section 3.3.  The foreground
contamination of this CMD is minimal, so the main sequence of Willman
1 is clearly distinguishable from foreground stars for $\sim$3
magnitudes below the turnoff. The mean photometric errors per 0.5
magnitude bin are overplotted. These errors include both measurement
and calibration uncertainties. In this Section, we use these
color-magnitude data to derive more robust estimates of the properties
of Willman 1 than possible with the shallower dataset used in
Paper I.  We also provide strong evidence that the properties of
Willman 1 are affected by tidal forces, and present the first evidence
for multi-directional tails and mass segregation.

\subsection{Age, metallicity and distance}

In Paper I, we compared the colors and magnitudes of the main sequence
turnoff (MSTO) and subgiant branch of Willman 1 to those of Palomar 5
and to those of \citet{girardi04} isochrones. We used these
comparisons to estimate Willman 1 to be an old, metal-poor population
at a distance of 45 $\pm$ 10 kpc.  We now compare the more precise
measurement of Willman 1's MSTO color ($g-r \sim 0.2$) and magnitude
($r \sim 21.9$) to those of the old globular clusters Pal 5, M5, M13,
and M15 as observed in SDSS to make a more robust estimate of Willman
1's distance and metallicity.  M15 is the most metal-poor of the four
comparison clusters, having [Fe/H] $\sim$ - 2.3. The other three have
more intermediate metallicities that all fall between -1.2 and -1.6
\citep{harrisGCcat}.  Although M15 has an MSTO color bluer than that
of Willman 1, the turnoff color of Willman 1 is consistent with that
of the other 3 clusters. This comparison suggests that Willman 1 has
-1.3 $\lesssim$ [Fe/H] $\lesssim$ -1.6.  This comparison is somewhat
undermined by the fact that turnoff color is sensitive to age as well
as metallicity.  The red giant branch color 1 - 2 mag brighter than
the turnoff is much less sensitive to age than the turnoff color.  Pal
5 ([Fe/H] = -1.2) is the only one of the four clusters with a
sub-giant branch that is redder than that of Willman 1, thus providing
additional support for our metallicity estimate. Comparing the
apparent magnitude of the MSTO of each of the four clusters with that
of Willman 1 leads to an average inferred distance of 38 kpc.
Assuming the absolute magnitude of Willman 1's MSTO lies in the range
of the four comparison clusters, and allowing for an uncertainty in
Willman 1's main sequence turnoff of $\pm$ 0.1 magnitudes gives a
total distance uncertainty of $\pm$ 7 kpc.

We also compare the stellar population of Willman 1 to a variety of
isochrones to obtain a check of our empirical determination. A
finely-spaced grid of old, metal-poor isochrones was constructed with
scaled-Solar abundances using an up-to-date version of the
\citet{chaboyer01} stellar evolution code.  Major improvements to the
code include the use of low-temperature opacities from
\citet{ferguson05} and the FreeEOS\footnote{Available from
http://feeeos.sourceforge.net/} equation of state \citep{irwin04}.
The isochrones were transformed to the observational plane using two
methods: first, a purely synthetic method based on PHOENIX model
fluxes \citep{hauschildt99a,hauschildt99b} and the throughput curves for
SDSS
filters\footnote{http://www.sdss.org/dr4/instruments/imager/index.html};
and second, a combination of semi-empirical $B$ and $V$ magnitudes
\citep{vandenberg03} and empirical equations relating $B$ and $V$ to
$g$ and $r$ \citep{smith02}.  Two color transformations were
employed to better understand the anticipated shortcomings of each
approach.  The comparisons of the CMD in Figure 1 with both color
transformations support our empirical finding that this object is old
(age $>$ 10 Gyr) and metal poor ([Fe/H] $\lesssim$ -1.3) but without
further observational constraints no stronger assertions can be made
with an isochrone analysis.

\begin{figure}
\plotone{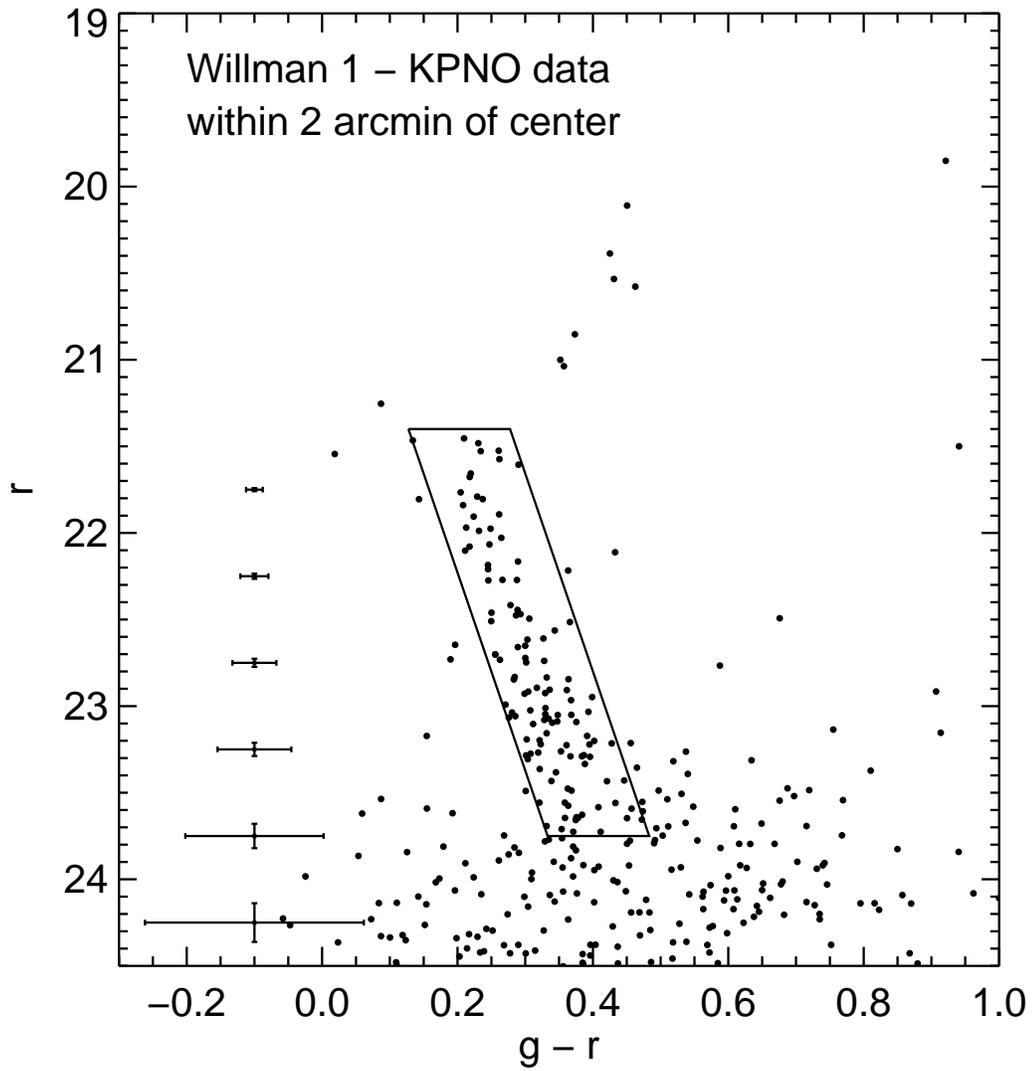}
 \caption{The color-magnitude diagram the central $2'$ of Willman
 1. The boxed region shows the color-magnitude range of stars used to
 produce the stellar density map in Figure 2.  A subgiant branch and
 main sequence are clearly visible in this plot. Average photometric
 uncertainties including both measurement and calibration errors are
 overplotted and only include errors of stars within the main-sequence
 box.}
\label{fig:W1CMD} 
\end{figure}

\placefigure{fig:W1CMD}



\subsection{Absolute Magnitude and Surface Brightness}

To improve the estimate of the total luminosity of Willman 1, we
compare the luminosity function (LF) of stars within its half-light
radius (r$_{1/2}$ = 1.9$'$; see \S3.3) with that of Palomar 5. We
selected Palomar 5 because it is old, moderately metal-poor, and
displays mass segregation like Willman 1 (see \S3.1 and 3.5). In
addition, its completeness corrected stellar LF is not strongly
affected by crowding, and is available in the
literature. \citet{koch04} presented the B band, foreground and
completeness corrected LF of stars within Pal 5's core radius
\citep[similar to its half-light radius;][]{harrisGCcat}.  We compute
the LF of stars within $r_{1/2}$ of Willman 1 that lie in the
main-sequence and sub-giant branch regions of the color magnitude
diagram.  We then determine the average field star LF in the 960
arcmin$^2$ region more distant than 11.4$'$ from the center
of Willman 1 and subtract it. In Table 1, we summarize these numbers.

\begin{table*}[htb]
\begin{center}
\caption{Stellar Luminosity Functions}
\begin{tabular}{lc c c c c c}
\colrule
\colrule 
& \multicolumn{2}{c}{$< r_{half}$} &
 \multicolumn{2}{c}{Central} &
\multicolumn{2}{c}{Tail} \\
{$r$} & N & N$_{corr}$ & N & N$_{corr}$ & N & N$_{corr}$ \\
\colrule
  21.50 -- 22.0 & 15.00 & 14.96 &  9.00 &  8.99 & 17.00 & 16.85\\
  22.00 -- 22.5 & 16.00 & 15.71 &  9.00 &  8.92 & 14.00 & 12.76\\
  22.50 -- 23.0 & 22.00 & 21.42 & 10.00 &  9.84 & 34.00 & 31.47\\
  23.00 -- 23.5 & 39.00 & 37.18 & 21.00 & 20.50 & 48.00 & 40.02\\
  23.50 -- 24.0 & 27.00 & 21.15 & 12.00 & 10.35 & 60.00 & 33.99\\
  24.00 -- 24.5 & 19.00 & 12.23 &  6.00 &  4.12 & 58.00 & 28.38\\
\colrule
\end{tabular}
\end{center}
 \tablecomments{The corrected values only include foreground
 subtraction; they do not include a completeness correction. The
 half-light, Central, and Tail areas are 11.34, 3.14, and 49.60 square
 arcminutes respectively.}
\end{table*}

To compare the LF to that of Pal 5, we used the transformations of
 \citet{smith02} to convert $g$ and $r$ to $B$.  Within its half-light
 radius, Pal 5 contains between 7 and 15$\times$ the number of stars
 that Willman 1 contains in each magnitude bin.  We only include bins
 brighter than 2 magnitudes below the MSTO so that completeness does
 not affect the comparison. We therefore scale the luminosity of Pal 5
 by the average LF ratio of 12 to obtain an absolute magnitude of
 $M_V\sim-2.5$ for Willman 1, consistent with (but a bit fainter than)
 the preliminary estimate in Paper I.  Assuming $M_V$ = -2.5 and
 $r_{1/2}$ = 1.9$'$, the average surface brightness within the
 half-light radius is 27.7 mag arcsec$^{-2}$. We obtain a minimum
 absolute magnitude of $M_V=-1.4$ by summing the luminosity of the
 foreground subtracted star counts of stars brighter than $r$=24.5 in
 the boxed region outlined in Figure 3, assuming a distance of 38 kpc.
 We thus estimate a generous uncertainty in the $M_V$ of Willman 1 as
 $\pm$ 1 mag.

\subsection{Spatial Extent and Tidal Radius}

 We now use the new data to show that Willman 1 has a half-light
radius consistent with that derived in Paper I, but has a larger
spatial extent than was possible to determine from the previous data.
In Figure 2, we plot the surface density profile of stars that lie
within the main sequence color-magnitude box overplotted on the CMD in
Figure 1. A mean stellar foreground of 0.44 stars arcmin$^{-2}$ (shown
by the dotted line) was determined as the average density of stars in
the region more than 11.4$'$ from the center of Willman 1
that lie in the appropriate color-magnitude box. The resulting profile
reaches the foreground level $\sim$10$'$ from the object center.
Integrating the profile shows that the radius containing half of the
stars is 1.9$'$. Allowing for an uncertainty of $\pm 0.3'$ in
the half-star radius and a distance uncertainty of $\pm$7 kpc yields a
physical size of 21$\pm$7 pc, very similar to that derived in Paper I.
We assume that this is a reasonable estimate of the half-light radius
$r_{1/2}$, since mass segregation (\S3.5) affects only the faintest
stars, which contribute little to the overall luminosity.

\placefigure{fig:profile}

To investigate the possibility that Willman 1 is tidally affected by
the Milky Way, we compare its spatial extent to its tidal
radius. Figure 2 shows that the total extent of Willman 1 stars is at
least 10$'$ from its center. We assume it is on a circular orbit,
treat the Milky Way as a point mass, and estimate the tidal radius as
r$_{tidal}$ = R$_{sat}$($M_{sat}/3M_{MW}$)$^{1/3}$, where $R_{sat}$ is
the distance between the satellite and the Milky Way and $M_{sat}$ is
the mass of the satellite (Equation 7-84, Binney \& Tremaine 1987). We
first assume that Willman 1 has a mass of 800 $ M_{\odot}$. This
mass was derived assuming $M_V = -2.5$ (see \S3.2) and a mass-to-light
ratio of 1, similar to that measured for low luminosity globular
clusters by \citet{mandushev91}. The resulting tidal radius is $r_{tidal}
\sim 2\arcmin$, comparable to the estimated $r_{1/2}$ and much smaller
than the total spatial extent of Willman 1.  Although the tidal radius
could be much larger if Willman 1 has a substantial dark matter
component, an increase in mass by a factor of 10 would still produce a
tidal radius significantly smaller than the total extent of its
stellar distribution.  Furthermore, if Willman 1's orbit is not
circular, its tidal radius during parts of its orbit is even smaller
than that derived here.  It thus is likely that Willman 1 is currently
strongly influenced by the tidal field of the Milky Way and is
unlikely that Willman 1 is currently in dynamical equilibrium. Palomar
5 is another example of such a stellar system.  \citet{dehnen04}
showed that the azimuthally averaged profile of its central and tail
regions suggests its stars extend to 107 pc, while its current tidal
radius is only $54$ pc currently and was even smaller (by a factor of
2) when Pal 5 was at perihelion.

\subsection{Multi-directional Stellar Tails}

To further investigate the possibility that Willman 1 is being tidally
  affected, we constructed a smoothed image of all stars that lie in
  the main sequence shown on the CMD in Figure 1.  Figure 3 shows this
  image, with contours corresponding to stellar surface densities that
  are 3 -- 20$\sigma$ above the field surface density.  The image was
  smoothed with an exponential filter of 0.3$'$ scale length. $\sigma$
  was calculated for the distribution of the smoothed surface
  densities of each image pixel, not including the center of the
  image.  The directions to the Galactic center and to the Ursa Major
  dwarf galaxy (UMa; only a few degrees away on the sky;
  \citealt{willman05b}) are overplotted.

The clumpy, tail-like morphology of Willman 1's isodensity contours
supports the idea that it is experiencing significant tidal
evolution. We obtained shallower observations of Willman 1 on the INT
2.5m in March 2005.  Although those observations are less sensitive to
the tails at very faint levels, they display the same 3 tail features
as the KPNO data. The presence of tails is not surprising, given
Willman 1's probable tidal radius.  However, such prominent
multi-directional features are unusual, particularly for a distant
object. Several GCs in the \citet{leon00} sample displayed possible
multi-directional features, including NGC 288 (d = 8.1 kpc).  However,
of the three most distant GCs in their sample (NGC 5694, 33 kpc; NGC
5824, 32 kpc; NGC 7492, 24.3 kpc), only NGC 5694 and 5824 show strong
evidence for tails, and neither show multi-directional tails.  Pal 5's
($d$ = 23 kpc) extensive tidal tails also do not appear
multi-directional.

How did a Milky Way companion at 40 kpc form such an unusual
 morphology?  It is not possible to determine without pursuing
 additional simulations and obtaining velocity information for this
 object.  However, we infer a couple of possibilities based only on
 some existing simulations. For example, Figure 6 of \citet{dehnen04}
 shows that a cluster on the derived orbit of Pal 5 is expected to
 display multi-directional tails when near apogalacticon.  The
 formation of streaky and S-shaped tidal features at apocenter is
 indeed a natural consequence of tidal tail evolution
 (\citealt{grillmair92}; C. Grillmair, private communication).  Future
 simulations may thus show that Willman 1's unusual morphology is
 evidence that it is near the apocenter of its orbit.

Another possibility is that the multiple tails of Willman 1 resulted
from gravitational shocking. Disk and bulge shocks are known to play a
significant role in the dynamical evolution of globular clusters on
orbits that bring them within a few kpc of the Galactic center
\citep{ostriker72,aguilar88,vesperini97,gnedin99}. \citet{combes99}
showed that multi-directional tails may be observed in objects that
have recently experienced gravitational shocking. Such objects display
tails perpendicular to the Galactic plane or along the Galactic
density gradient.  Willman 1's significant distance from the Milky Way
makes gravitational shocking from the Milky Way itself an unlikely
explanation for the multi-directionality.  However, the morphology of
Willman 1 could be a combination of the tidal effects of the Milky Way
and of interaction with another outer halo object. \citet{knebe05}
found that satellite-satellite interactions account for $\sim$ 30\% of
total satellite mass loss, although penetrating encounters between
satellites are relatively rare.  Furthermore, preliminary simulations
of globular cluster evolution in a Milky Way-like halo have produced
clusters with unusual morphologies when the simulated halo includes
Cold Dark Matter sub-halos (L. Mayer, private communication).  In a
future paper, we will pursue a variety of possibilities using
numerical simulations informed by velocity data.

\subsection{Evidence for Mass Segregation}

 A relative overabundance of low mass stars has been observed in the
 outskirts of many Milky Way globular clusters
 \citep{king95,lee04}. There are thought to be two primary sources of
 this mass segregation in clusters: two-body relaxation
 \citep{binney87} and a primordial spatial variation of the initial
 mass function (IMF).  Two body interactions transfer energy from more
 to less massive stars, placing the less massive stars on orbits with
 higher average radius.  Dynamical mass segregation occurs on the
 relaxation timescale of a system \citep{binney87}. However, star
 clusters that are younger than the relaxation timescale have also
 been observed to display mass segregation. This is evidence for the
 existence of primordial mass segregation, whereby the IMF was
 different at different cluster locations
 \citep{bonnell98,sirianni02}.

Tidal stripping exacerbates mass segregation because stars that lie at
large radii are preferentially stripped. One thus expects that the
tidal tails of a mass segregated object will be rich in low mass stars
relative to the inner regions of the object.  For example, mass
segregation has been established in both NGC 288
\citep{bellazzini02} and Palomar 5, low central density clusters that
have been shown to be affected by disk shocking. As \citet{koch04}
explain, unless Palomar 5 used to be much more centrally concentrated
it is unlikely that the mass segregation was primarily due to two-body
relaxation.  Its mass segregation may thus be a combination of
primordial and evolutionary effects.

To look for evidence of mass segregation in Willman 1, we compare the
stellar luminosity functions of the central and tail regions. In
Figure 4, we plot the central and tail LFs plus an arbitrary
constant. The error bars include Poisson error in both the number of
object stars and in the field subtraction. The ``central'' region LF
includes everything within 1$'$ of the object center (area =
3.14$'^2$) and the ``tail'' region includes everything within the
boxes outlined on Figure 3 but outside of 1$'$ (area = 46.46$'^2$).
These LFs are summarized in Table 1.  

A KS test shows that there is a 68\% chance that the two stellar LFs
were drawn from different populations.  At $\sim2\sigma$ significance,
the central field contains fewer faint, low mass stars than the
tail at faint magnitudes. The central LF decreases by 50$\%$ in the
bin 2 magnitudes fainter than the MSTO ($23.5 < r < 24.0$), whereas
the LF of stars in the tail remains nearly constant. This difference
between the central and tail LFs cannot be explained by completeness,
because our data are not confusion limited (see \S2.1).  The same is
true for stars between $24.0 < r < 24.5$.  A similar decrease in the
relative number of faint stars in the center and in the tails is also
seen in Pal 5.  However in Pal 5 the lack of faint stars does not
become apparent until 3 magnitudes fainter than its MSTO. In contrast,
the lack of faint stars in the center of Willman 1 appears at only 2
magnitudes fainter than the MSTO. This difference suggests that mass
segregation may be affecting higher mass stars in Willman 1 than in
Pal 5. Deeper imaging will confirm whether this trend continues at
fainter magnitudes.  We briefly discuss the possible implications of
the observed mass segregation is \S4.

\section{Discussion}

In this paper, we have used deep, wide-field imaging to improve
estimates of the distance, metallicity, absolute magnitude, and size
of Willman 1. This object has a total luminosity that is similar to
that of the least luminous globular clusters.  Half-light radius is
often used to characterize the size of globular clusters and dwarf
galaxies because it does not easily evolve in response to tidal or
internal dynamical evolution and is thus a robust reflection of the
object's size at formation. Paper I showed that Willman 1 has a
half-light size larger than one might expect for a low luminosity
globular cluster, raising the question whether Willman 1 may actually
be an extreme dwarf galaxy. In this paper, we have shown evidence that
Willman 1 displays: a spatial extent that may exceed its tidal radius,
multi-directional tidal tails and possible mass segregation. These
results suggest that Willman 1's large half-light size may be an
indication that it is on the verge of disruption rather than an
indication that its formation was more similar to that of dwarf
galaxies than that of globular clusters.

These new results provide tantalizing hints to the nature of this
ambiguous object. The large spatial extent of Willman 1 relative to
its tidal radius shows that it is unlikely to be in dynamical
equilibrium and that its evolution is being strongly affected by the
tidal influence of the Milky Way, even if dark matter makes up 90\% of
the total mass of the system.  As discussed in \S3.4, the presence of
multiple tidal tails raises several possibilities for the orbit and
history of Willman 1. Its multiple tails make Willman 1 a very
interesting object to compare with numerical models of substructure
evolution in a lumpy galaxy halo, which we will do in a subsequent
paper.

The mass segregation observed in Willman 1's stars has not been
previously observed in dwarf galaxies, but rather has only been seen
in globular clusters.  Although some star clusters provide evidence
for primordial mass segregation, most GCs are thought to be mass
segregated due to dynamical effects.  Because the central stellar
densities of known, nearby dwarf galaxies are much lower than those of
globular clusters, their relaxation timescales are too long for them
to exhibit dynamical mass segregation.  Willman 1's current central
density is the same as that of the Milky Way dSphs, but may have been
far higher in the past if it has since undergone gravitational
shocking.  Willman 1 would likely have formed in a similar way to
known globular clusters if it had formed with a high enough central
density to undergo significant dynamical mass segregation.  The
situation is less clear if the mass segregation was primordial in
origin.

Due to the small stellar mass of Willman 1, these results are still
consistent with the possibility suggested by
\citet{willman05a} that it may have formed within a low mass dark
matter halo. Although this scenario is extreme, we discuss it here to
underscore the increasingly ambiguous distinction between these two
classes of objects \citep{huxor05,hasegan05}. The stellar mass of
Willman 1 is so small that its spatial extent exceeds it tidal radius
even if it is dark matter dominated.  Furthermore, it may be difficult
for an object as low luminosity as Willman 1 to host multi-directional
tails, unless it also hosts a dark matter halo.  How long could a 800
$M_{\odot}$ object appear to have a coherent structure after
experiencing an event that causes prominent multi-directional tails?
Preliminary simulations of globular clusters orbiting within a CDM
galaxy halo suggest that very low mass globular clusters are unlikely
to remain intact in the event of a substantial tidal event (L. Mayer,
private communication).  Upcoming multi-object spectroscopy of Willman
1 stars may be able to determine if even its central region is
gravitationally bound.

Although the results presented in this paper have shed considerable
new light on the nature of Willman 1, the present data still only hint
at a range of possibilities for its progenitor. Willman 1's current
life expectancy as an object with an order of magnitude fewer stars
than Pal 5 may be so short (e.g. few Myr) so as not to expect to ever
observe such an object but for a great coincidence.  Perhaps many more
similar objects used to exist but have been destroyed beyond
detection. The extremely small number of stars in Willman 1, and the
fact that it contains few, if any, horizontal branch or red giant
branch stars exacerbate the difficulty of determining its fundamental
properties. Although it contains only a small number of stars bright
enough to obtain precise radial velocities for, it is possible that a
detailed kinematic study will shed light on Willman 1's mass-to-light
ratio and dynamical state. In the future, a combination of deeper
photometry, kinematic data, and numerical modeling will hopefully
unravel the mystery of Willman 1.


\acknowledgements
 
We acknowledge Carl Grillmair for interesting and informative
discussions via email. We thank Anil Seth for discussions that
significantly contributed to the stellar photometry and completeness
testing. We acknowledge the contributions of Scott Burles, Phil
Marshall, and Sam Roweis in the development of the software used to
perform the astrometry. MM, DWH, and MRB are partially supported by
NASA (LTSA grant NAG5-11669) and the NSF (grant AST-0428465). MM and
DWH are grateful to the MIT Kavli Institute for Astrophysics and Space
Research for hospitality during the period of this research. JJD. and
AAW were partially supported through National Science Foundation grant
CAREER AST 02-38683 and the Alfred P. Sloan Foundation. DMD recognizes
support by the Spanish Ministry of Education and Science (Ramon y
Cajal contract and research project AYA 2001-3939-C03-01)


\ifsubmode\else
\baselineskip=10pt
\fi


\bibliographystyle{apj}
\bibliography{../master}

\clearpage

\begin{figure}
\plotone{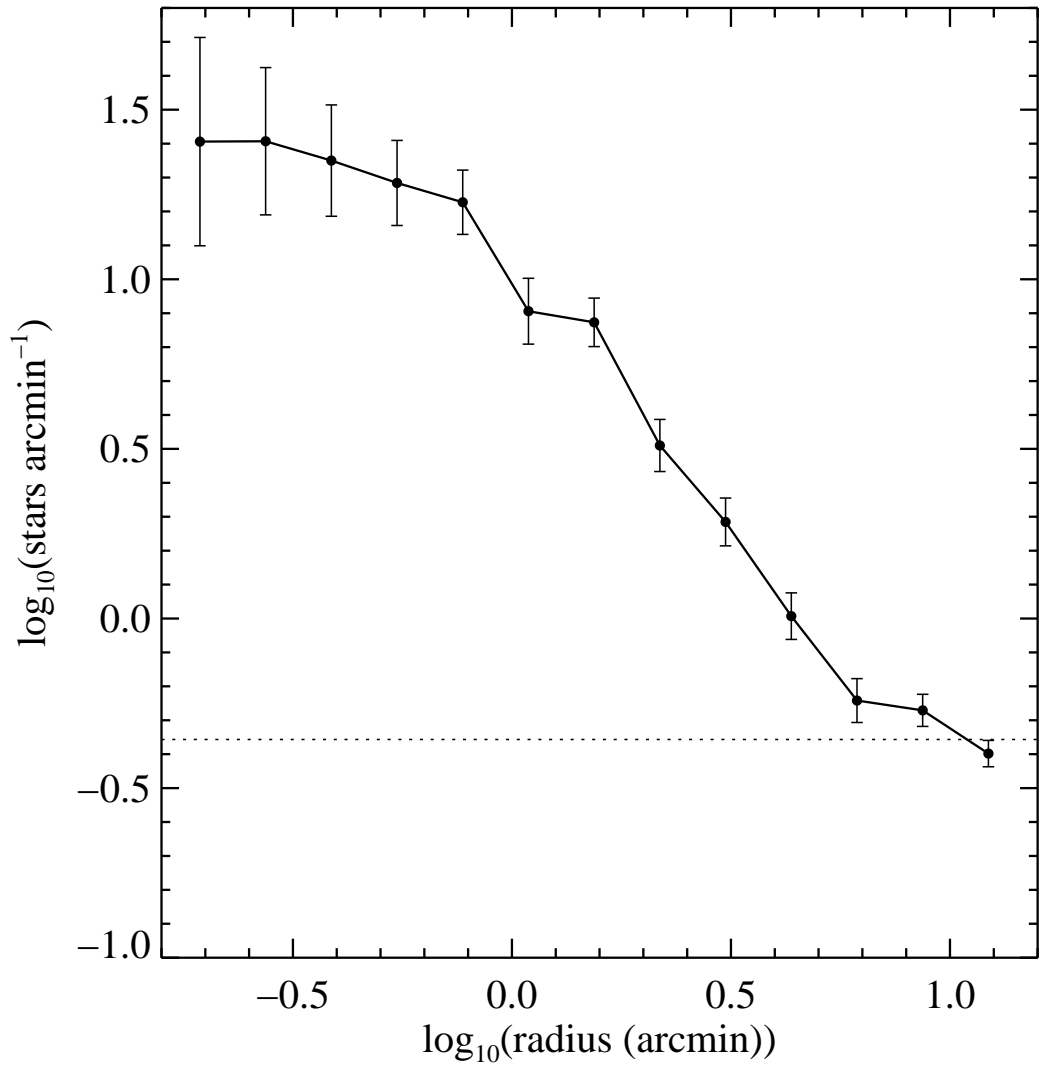}
 \caption{The azimuthally averaged radial profile of stars brighter
than $r$ = 23.75 that lie within the main-sequence box overplotted on
Figure 1.  The foreground has not been subtracted. The dotted line
shows the average number density of main sequence stars more than 5
half-light radii from the center of Willman 1 and brighter than $r$ =
23.75. Willman 1 stars extend to at least $10'$ from its center. The
error bars were calculated assuming Poisson statistics.}
\label{fig:logprofile}
\end{figure}

\begin{figure}
\plotone{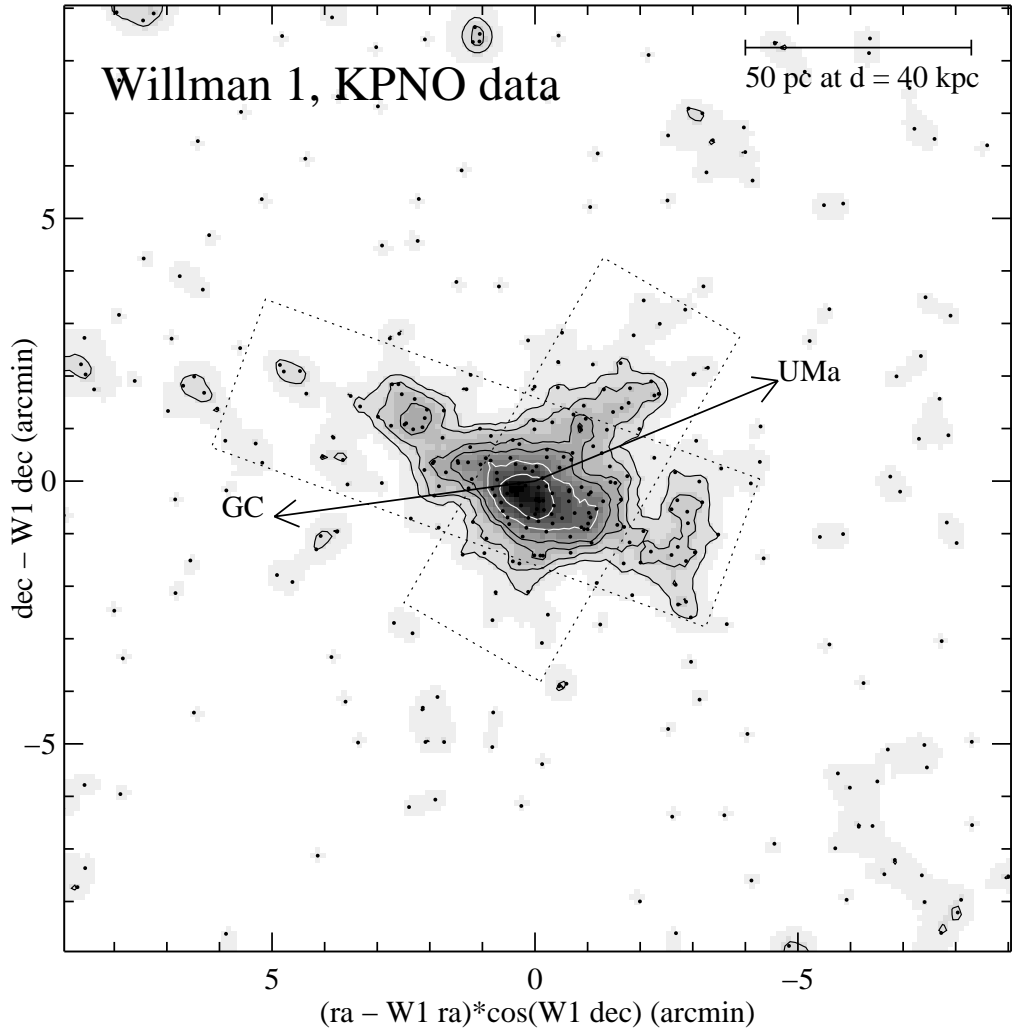}
\caption{Smoothed spatial
distribution of stars brighter than $r$ = 23.75 that lie within the
main-sequence box overplotted on Figure 1. The position of each star
used to create this map is shown by a dot in the Figure. The contours
correspond to stellar surface densities that are 3, 5, 8, 10, 15, and
20$\sigma$ above the field surface density. The 15 and 20$\sigma$
contours are plotted in white. When computing the stellar luminosity
function, we define the ``tail'' as the region outlined by the dotted
boxes, but not including the central 1$'$.  Note the prominent
multi-directional tidal features. The directions to the Ursa Major
dwarf galaxy and to the Galactic Center are overplotted.}
\label{fig:W1distr}
\end{figure}

 \begin{figure} 
\plotone{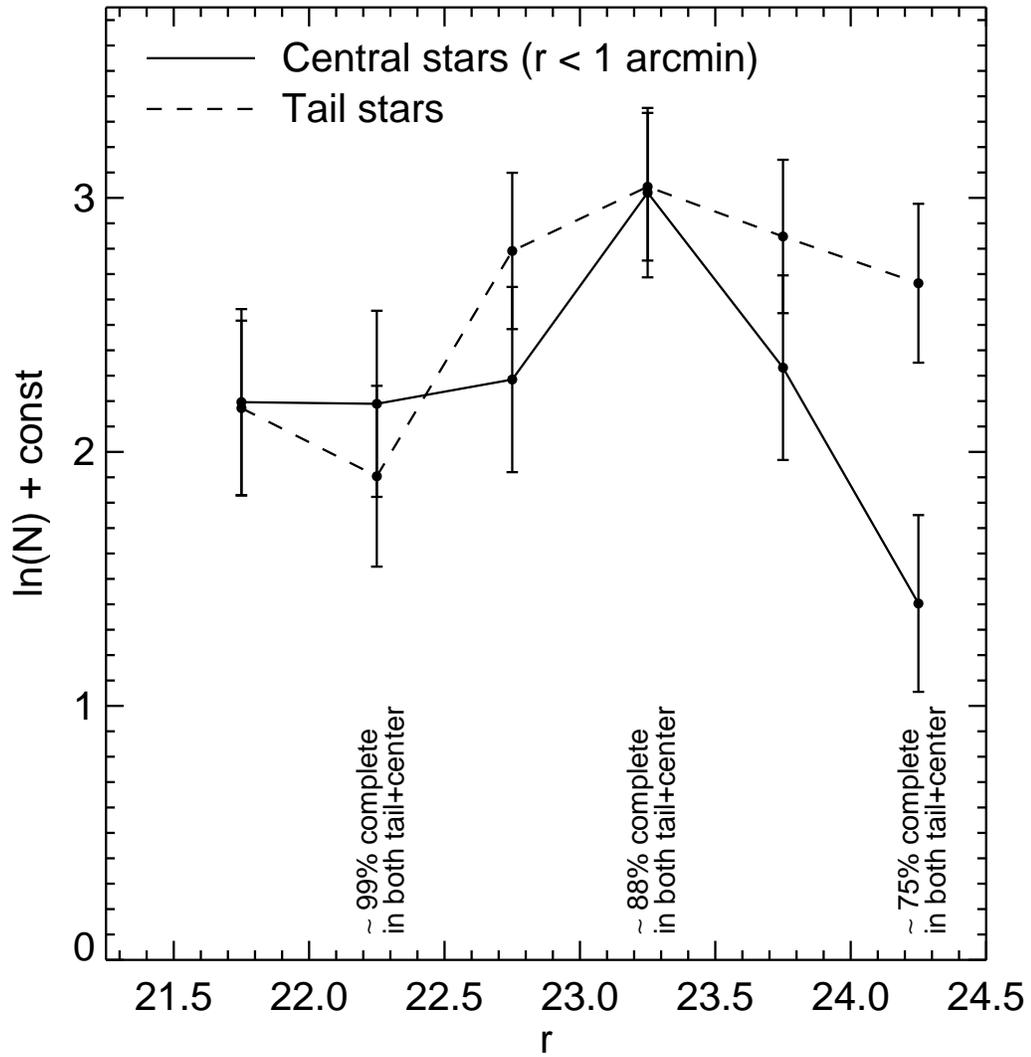} 
 \caption{The natural log of the foreground corrected luminosity
functions of stars within 1$'$ of Willman 1 and of stars within the
tail regions outlined on Figure 2, but not including stars in the
central 1$'$.  The tail star LF is offset by a constant to facilitate
comparison with the central star LF.  The bins are 0.5 magnitude wide
and bin centers are plotted.  The MSTO is near $r = 21.5$.  Errors
were calculated assuming Poisson fluctuations in the number of object
and foreground stars. The tidal tails contain more low luminosity
stars than the center at $> 2\sigma$ significance.}
\label{fig:LF} 

\end{figure} 

\end{document}